\documentclass[prd,preprint,showpacs,eqsecnum,floatfix]{revtex4}

\usepackage{amsfonts}
\usepackage{amssymb}
\usepackage{graphicx}
\usepackage{pstricks}

\newcommand{\Tr}{{\rm Tr}}

\newcommand{\BOX}{\hbox {$\sqcap$ \kern -1em $\sqcup$}}

\newcommand{\be}{\begin{equation}}
\newcommand{\ee}{\end{equation}}
\newcommand{\ba}{\begin{eqnarray}}
\newcommand{\ea}{\end{eqnarray}}
\newcommand{\ban}{\begin{eqnarray*}}
\newcommand{\bea}{\begin{eqnarray}}
\newcommand{\eea}{\end{eqnarray}}
\newcommand{\ean}{\end{eqnarray*}}
\newcommand{\barr}{\begin{array}}
\newcommand{\earr}{\end{array}}

\begin{document}

\title{Quantum Decoherence and Neutrino Data}

\author{Gabriela~Barenboim}
\affiliation{Departamento de F\'isica Te\'orica and IFIC, Centro Mixto,
Universidad de Valencia-CSIC, E-46100, Burjassot (Valencia), Spain.}
\author{Nick~E.~Mavromatos}
\author{Sarben~Sarkar}
\author{Alison ~Waldron-Lauda}
\affiliation{King's College London, University of London,
Department of Physics, Strand WC2R 2LS, London, U.K.}

\begin{abstract}

In this work we perform
global fits of microscopic decoherence models
of neutrinos to all available current  data, including LSND and
KamLAND spectral distortion results. In  previous works on related issues the models used were supposed to explain
LSND results by means of quantum gravity induced decoherence. However those models were purely phenomenological
without any underlying microscopic basis. It is one of 
the main purposes of this article to use detailed microscopic
decoherence models with complete positivity, to fit the data.The decoherence in these models has contributions  not
only from stochastic quantum gravity vacua operating 
as a medium, but also from conventional uncertainties in the energy
of the (anti)neutrino beam. All these 
contributions lead to oscillation-length 
independent damping factors modulating the
oscillatory terms from which one
 obtains an excellent fit to all available neutrino data, 
including LSND and Kamland spectral distortion. The fit is much
 superior to all earlier ones.
It appears that the results of the  
fit are most naturally interpreted as 
corresponding to conventional energy uncertainties. This
represents a radical departure from previous analyses 
where the neutrino data (including LSND but not Kamland spectral
distortion) were regarded as evidence for quantum gravity decoherence.

\end{abstract}
\preprint{FTUV-06-02xx}

\maketitle

\section{Introduction}

The theory of Quantum Gravity (QG) is still elusive.
In some theoretical models, the phenomenon of space-time `foam',
invoked by J.A. Wheeler~\cite{wheeler}, may be in place;
according to this picture the singular microscopic fluctuations
of the metric, give the ground state of QG the structure of a `stochastic medium'. The medium
has the profound effect of leading to decoherence of quantum matter as it propagates.This may have experimentally observable
consequences in principle~\cite{poland}. 

One of the basic effects of decoherence is the presence of damping factors 
in front of the oscillatory terms.
However, one should be very careful 
when interpreting decoherence effects, if observed in an experiment,
because ordinary matter can easily `fake' decoherence effects, especially 
the damping exponents~\cite{ohlsson,ohlsson2}. For instance, 
uncertainties in the energy of a neutrino beam~\cite{ohlsson}, which 
are associated with ordinary physics, and have nothing to do 
with `fuzziness' of space time, do reproduce a damping exponent 
similar to that encountered in Lindblad decoherence models~\cite{lindblad}. 
Of course, stochastic quantum gravity effects can induce such uncertainties
in the energy beam, and hence contribute to the damping exponent 
themselves~\cite{poland}, but such effects are usually subleading. 
Thus, one should know the energy of the beam with high precision
in order to eliminate `fake' decoherence effects and probe quantum gravity 
effects sufficiently well. 

In ref.~\cite{barenboim2}, henceforth referred to as I,
we have attempted to fit the available neutrino data,
including LSND results~\cite{lsnd}, using phenomenological
decoherent models with mixing in all three generations of neutrinos.
Such fits extended earlier similar
attempts to study decoherence with two-generation neutrino models~\cite{lisi}.
In I it was seriously entertained that
the decoherence might be attributable to environmental
entanglement with the quantum gravity foamy vacuum,
and could be distinguished
from ordinary-matter-induced decoherence~\cite{poland}.

A simplified model, of Lindblad type~\cite{lindblad}
has been used for the fit, following earlier work in \cite{gago}.
The model of \cite{gago} involved a phenomenological diagonal decoherence matrix,
and in the fit of I, decoherence was assumed to be dominant
{\it only} in the antineutrino sector. This assumption was made in order to fit the LSND
results~\cite{lsnd} pointing to
significant ${\overline \nu}_\mu \leftrightarrow {\overline \nu}_e$
oscillations,  but no significant evidence of
oscillations in the particle sector.
In this way a fit was made to a three generation model
with the LSND ``anomalous'' result, without introducing a sterile neutrino.
The possibility of strong CPT violation in the decoherence sector, allowed for
an equality of neutrino and antineutrino mass differences in agreement with atmospheric and solar neutrino data.

The particular choice of \cite{barenboim2}, which yielded an extremely good fit to all available neutrino data involved mixed energy dependence
for the (antineutrino-sector) decoherence coefficients, some of
which were proportional to the neutrino energies $E$, while the rest
were inversely proportional to it, $\propto 1/E$. In I, the
coefficients proportional to $1/E$ were interpreted as describing
ordinary matter effects, whilst those proportional to $E$ were
assumed to correspond to {\it genuine} quantum-gravity effects.
The latter increase with the energy of the the (anti)neutrino is consistent with a larger back reaction effect on quantum space-time
and hence with a larger decoherence.

The strong difference assumed in I between the decoherence
coefficients of the particle and antiparticle sectors, although not
incompatible with a breakdown of CPT at a fundamental
level~\cite{poland}, appears at first sight somewhat curious, and in
fact is unlike any other case of decoherence in other sensitive
particle probes, like neutral mesons, examined in the
past~\cite{lopez}. There, the oscillations between particle and
antiparticle sectors, necessitate a common decoherence environment
between mesons and antimesons. If one accepts the Universality of
gravity, then, the sample point of I seems incompatible with this
property. Moreover, there are two more problematic points of the fit in I,which were already discussed in that reference.
The first point concerns the complete positivity of the model. In \cite{gago} the {\it ad hoc} diagonal form of the decoherence
matrix, used in I, was postulated , without a discussion of the necessary conditions required in the Lindblad approach to guarantee complete positivity.
Indeed the particular choice of the decoherence parameters of I, did {\it not} lead to
positive definite probabilities for the {\it entire}
regime of the parameter space of the model, although the
probabilities were positive definite for the portion of
the parameter space appropriate for the
various neutrino experiments  used for the fit. Specifically,
it was found in I, that with the particular choice of the
decoherence parameters in the (antineutrino sector),
one obtains positive-definite transition probabilities
for energies restricted to $E > {\cal O}(1~{\rm MeV})$.
The second, and more important point, is that the choices of decoherence parameters of I
were  good for all the neutrino experiments available at the time,
but {\it unfortunately} it could not reproduce the spectral distortion
observed by the KamLAND experiment~\cite{kamland}, whose first results
came out simultaneously with the results of I.

The aim of this article is two fold.
One is to rectify the above points, 
and present a novel fit, using as in I the simplified three-generation
Lindblad model of decoherence of \cite{gago}, but crucially
amended so as to respect the general conditions among the coefficients necessary to guarantee complete positivity in the {\it entire} parameter space.
We shall show below, that it is possible to find such 
a consistent Lindblad model
consistently, for which the fit to all available neutrino data is
excellent, including the spectral distortion seen by KamLAND, and
the LSND results~\cite{lsnd}.
This substantially extended decoherence model (in comparison to I), constitutes therefore the first mathematically
consistent three-generation neutrino decoherence model of Lindblad
type which fits all the available data, including spectral distortion
seen by KamLAND. 

The second and more significant aim was to give 
a microscopic and physically motivated model which would fit 
into the general scheme of the linear Lindblad decoherence. 
In this way the constraints obtained from the phenomenological fit 
can be examined to check consistency with values that can be 
deemed reasonable for phenomena originating from quantum gravity. 
Such a comparison has not been done before and leads to a major shift 
in our views concerning decoherence due to quantum gravity.
In fact, as we shall argue below, the most natural  explanation of 
the fit seems to be provided by 
energy uncertainties
in the (anti)neutrino beam, due to conventional physics.
Several microscopic quantum space-time 
models, that we have examined in this work, yield too small effects
to reproduce the result of the fit. 

In the present work, 
the decoherence parameters in the model are {\it assumed} to be the
same in {\it both neutrino and antineutrino sectors},
consistent with the above-mentioned universal property of a
quantum-gravity environment. In this sense, we assume that the
LSND result is correct in both channels, although their observed
excess of ${\overline \nu}_e$ events is not corroborated (at the
same level at least) in the neutrino channel.
This is to be
contrasted with the approach of I, where following ~\cite{bl},
only the evidence in the
antineutrino sector was considered.

The structure of this article is the following: in section 2 we
review the basic theory of Lindblad decoherence, and specify the
conditions for complete positivity in the type of model for
decoherence used in \cite{gago} and in I. We examine the limitations
imposed on the parameter space of the model in I in order to
guarantee complete positivity, and then we construct a modified model,
in which one obtains positive definite transition probabilities
for the entire regime of the parameter space. The decoherence implies
exponential damping with time (oscillation length), which
violates microscopic time irreversibility, irrespective of CP
properties of neutrinos, and hence CPT violation. In order to agree with the experimental results of KamLAND on spectral distortions~\cite{kamland} we require such
exponential damping factors to imply a modulation
in front of some of the oscillatory terms giving rise to a modification in the (survival)
transition probabilities of order per mil. We obtain
in section 3 stringent constraints on the exponents of the damping
factors. The sample point that fits all available data, including LSND,is discussed in detail in section 3. This model-point
 corresponds to an exponent of the decoherent damping factors
which is independent of the oscillation length. An attempt to
explain such a result in terms of microscopic models of stochastic
space time foam is given in section 4.
However, as we show there, explanations based on conventional physics  
such as the uncertainties in the (anti)neutrino beam energy, 
are definitely much more plausible. The present data when 
interpreted in terms of a microscopic model make quantum gravity 
an unlikely candidate for the origin of decoherence (claimed to 
be observed in our fits).
Finally, conclusions and outlook are presented in section 5.

\section{Lindblad Decoherence and Transition Probabilities: a review}

In this section we present the details of the
calculation for transition probabilities of
three generations of neutrinos, where complete positivity is maintained
within the (linear) Lindblad approach~\cite{lindblad}.
In this framework the general evolution equation of the $\rho$ density matrix, representing a
(spinless) neutrino state reads :
\ba
\partial_t \rho = L[\rho ]
\label{linearlindblad} \ea where there are conditions on the
decoherence contribution to $L$ which guarantee complete positivity
of the probabilities as they evolve in time. The spin of the
neutrino will not play an important r\^ole in constraining the
decoherence sector by comparing with experimental data, and hence we
shall present a formalism based on scalar particles. Detailed
studies of Dirac and spinless neutrinos have been performed in
\cite{msw2}; as explained there the inclusion of spin does not
affect qualitatively the main decoherence effects which is the
damping of oscillation probabilities. In section 4, where we attempt
to interpret the time (i.e. oscillation length) dependence and order
of magnitude of the decoherence parameters, we shall present a more
detailed discussion on the results of \cite{msw2}.

With the above in mind, we commence our analysis with a theorem due to
Gorini, Kossakaowski and Sudarshan ~\cite{gorini} on the structure of $L$, the generator of a quantum dynamical semi-group~\cite{lindblad,gorini}.
For a non-negative matrix $c_{kl}$ (i.e. a matrix
with non-negative eigenvalues) such a generator is given by
 \ba
L[\rho]=-i[H,\rho]+\frac{1}{2}\sum_{k,l}c_{kl}\left([F_k\rho,F^{\dagger}_l]
    + [F_k,\rho F^{\dagger}_l]\right),
 \ea
where $H=H^{\dagger}$ is a hermitian Hamiltonian,
$\{F_k,k=0,...,n^2-1\}$ is a basis in $M_n(\textbf{C})$ such that
$F_0=\frac{1}{\sqrt{n}}I_n$, Tr$(F_k)=0\forall k\neq 0$ and
Tr$(F^{\dagger}_i F_{j})=\delta_{ij}$. In our application we can take
$
F_i  = \frac{{\Lambda _i }}{2}
$
 (where ${\Lambda _i}$ are
the Gell-Mann matrices) and satisfy the Lie algebra $[F_i,F_j]=i
\sum_k f_{ijk}F_{k}, (i=1,...8)$ $f_{ijk}$ being the standard
structure constants, antisymmetric in all indices.

Without a microscopic model, in the three generation case, the precise physical
significance of the decoherence matrix cannot be fully understood.
In this work we shall consider the simplified case in which
the matrix $C \equiv (c_{kl})$ is {\it assumed} to be
of the form
 \ba
  C =   \left(%
\begin{array}{cccccccc}
  c_{11} & 0 & 0 & 0 & 0 & 0 & 0 & 0 \\
  0 & c_{22} & 0 & 0 & 0 & 0 & 0 & 0 \\
  0 & 0 & c_{33} & 0 & 0 & 0 & 0 & c_{38} \\
  0 & 0 & 0 & c_{44} & 0 & 0 & 0 & 0 \\
  0 & 0 & 0 & 0 & c_{55} & 0 & 0 & 0 \\
  0 & 0 & 0 & 0 & 0 & c_{66} & 0 & 0 \\
  0 & 0 & 0 & 0 & 0 & 0 & c_{77} & 0 \\
  0 & 0 & c_{38} & 0 & 0 & 0 & 0 & c_{88} \\
\end{array}%
\right)
\label{cmatrix}
 \ea
As stated above, positivity can be guaranteed if and only if
the matrix $C$ is positive
and hence has non-negative eigenvalues.  We have also taken $C$
to be symmetric. A similar simplification has been used in
\cite{gago} and in I, to yield an economic decoherence
model which can be compared with experimental data. However, as discussed in detail in \cite{msw2}, such special choices can be realised for models of
the propagation of neutrinos in  models of stochastically
fluctuating environments~\cite{loreti}, where the decoherence term
corresponds to an appropriate double commutator involving operators that entangle with the environment.The quantum-gravity space time
foam may in principle behave as one such stochastic
environment~\cite{poland,msw2,lopez}, and it is this point of view,
that we will critically examine in this work. In section 4,  we shall discuss the viability of the interpretation of the
fit in terms of such microscopic models of space-time
foam.

Since the $F_i$ are Hermitian, we can rewrite the
expression for $L[\rho]$ as
 \ba
    L[\rho] \equiv -i[H,\rho] + {\cal D}[\rho] =
   -i[H,\rho]+\frac{1}{2}\sum_{k,l}c_{kl}
    \left([F_k\rho,F_l]+ [F_k,\rho F_l]\right)
 \ea
After standard manipulations,
we may write the non-Hamiltonian decoherence part ${\cal D}[\rho ]$
as
\ba
  {\cal D}[\rho ] =
\frac{1}{4}\left([F_k,[\rho,F_l]]+\{F_k,[\rho,F_l]\}-[F_l,[F_k,\rho]]
    +\{F_l[F_k,\rho]\}\right)+\frac{1}{2}\{\rho,[F_k,F_l]\}
\ea
On using the expansion $\rho=\sum_i \rho_i F_i$,
this expression can be written
 \ba
  {\cal D}[\rho]= \frac{\rho_i c_{kl}}{4}\left(-f_{ilm}f_{kmj}F_j+if_{ilm}(\frac{1}{3}\delta_{km}
    +\frac{1}{2}d_{kmj}F_j)+f_{kim}f_{lmj}F_j
    \right.
  \nonumber   \\ \left.+if_{kim}(\frac{1}{3}\delta_{lm}+\frac{1}{2}f_{lmj}F_j)
    +i2f_{klm}(\frac{1}{3}\delta_{im}+\frac{1}{2}d_{imj}F_j)\right)
 \ea
We note that the only terms which contribute are $\frac{\rho_i
c_{kl}}{4}\left(-f_{ilm}f_{kmj}
    +f_{kim}f_{lmj}\right)F_j$.

We follow the basic notation~\cite{lindblad,gago}
and express the time evolution of the
density matrix as
 \ba
    \dot{\rho}_k=\sum_j(\sum_i h_i f_{ijk}+{\cal D}_{kj})\rho_j=\sum_j
    M_{kj}\rho_j
 \ea
where we have
 \ba
    {\cal D}_{ij}=\sum_{k,l,m} \frac{c_{kl}}{4}\left(-f_{ilm}f_{kmj}
    +f_{kim}f_{lmj}\right).
 \ea
Using the values of the structure constants $f_{ijk}$
of the SU(3) group, appropriate
to the three generation case being examined here,
we arrive at:
 \ba
    {\cal D}_{11}&=& -\frac{1}{2}\left( c_{22}+c_{33}+\frac{1}{4}
    (c_{44}+c_{55}+c_{66}+c_{77})\right) \nonumber
    \\ {\cal D}_{22}&=& -\frac{1}{2}\left( c_{11}+c_{33}+\frac{1}{4}
    (c_{44}+c_{55}+c_{66}+c_{77})\right) \nonumber
    \\ {\cal D}_{33}&=& -\frac{1}{2}\left( c_{11}+c_{22}+\frac{1}{4}
    (c_{44}+c_{55}+c_{66}+c_{77})\right) \nonumber
    \\ {\cal D}_{44}&=& -\frac{1}{2}\left( c_{55}+\frac{1}{4}
    (c_{11}+c_{22}+c_{33}+c_{66}+c_{77}+3c_{88})+\frac{\sqrt{3}}{2}
    c_{38} \right) \nonumber
    \\ {\cal D}_{55}&=& -\frac{1}{2}\left( c_{44}+\frac{1}{4}(c_{11}+c_{22}+c_{33}+
    c_{66}+c_{77}+3c_{88})+\frac{\sqrt{3}}{2}
    c_{38} \right) \nonumber
    \\ {\cal D}_{66}&=& -\frac{1}{2}\left( c_{77}+\frac{1}{4}(c_{11}+c_{22}+c_{33}+
    c_{44}+c_{55}+3c_{88})-\frac{\sqrt{3}}{2}
    c_{38} \right) \nonumber
    \\ {\cal D}_{77}&=& -\frac{1}{2}\left( c_{66}+\frac{1}{4}(c_{11}+c_{22}+c_{33}+
    c_{44}+c_{55}+3c_{88})-\frac{\sqrt{3}}{2}
    c_{38} \right) \nonumber
    \\ {\cal D}_{88}&=& - \frac{3}{8}(
    c_{44}+c_{55}+c_{66}+c_{77}) \nonumber
    \\{\cal D}_{83}&=& {\cal D}_{38}=- \frac{\sqrt{3}}{8}(
    c_{44}+c_{55}-c_{66}-c_{77})
\label{clrelation}
 \ea
or conversely,
\ba
   \nonumber c_{{11}}&=&\frac{1}{3}{\cal D}_{{88}}+{\cal D}_{{11}}-{\cal D}_{{2 2}}
   -{\cal D}_{{33}}
    \\\nonumber c_{{22}}&=&-{\cal D}_{{11}}+\frac{1}{3}{\cal D}_{{88 }}+{\cal D}_{{22}}-{\cal D}_{{33}}
    \\\nonumber c_{{33}}&=&\frac{1}{3}{\cal D}_{{88}}-{\cal D}_{{11}}-{\cal D}_{{22}}+{\cal D}_{{33}}
    \\\nonumber c_{{44}}&=&-{\cal D}_{{ 55}}+{\cal D}_{{44}}-\frac{2}{\sqrt{3}}
    {\cal D}_{{38}}-\frac{2}{3}{\cal D}_{{ 88}}
    \\ c_{{55}}&=&{\cal D}_{{55}}-{\cal D}_{{44}}-\frac{2}{3} \sqrt {3}{\cal D}_{{38}}
    -\frac{2}{3}{\cal D}_{{88}}
     \\\nonumber c_{{6 6}}&=&-{\cal D}_{{77}}+{\cal D}_{{66}}+\frac{2}{\sqrt{3}}
   {\cal D}_{{38}} -\frac{2}{3}{\cal D}_{{88}}
     \\ \nonumber c_{{77}}&=&{\cal D}_{ {77}}-{\cal D}_{{66}}+\frac{2}{\sqrt {3}}
     {\cal D}_{{38}}-\frac{2}{3}{\cal D}_{{88}}
     \\ \nonumber c_{{88}}&=&-\frac{2}{3}\,{\cal D}_{{55}}-\frac{2}{3}{\cal D}_
    {{77}}-\frac{2}{3}{\cal D}_{{66}}-\frac{2}{3}{\cal D}_{{44}}+{\cal D}_{{88}}
    +\frac{1}{3}{\cal D}_{{11}}+\frac{1}{3}{\cal D}_{{22}}+\frac{1}{3}{\cal D}_{{33}}
    \\ \nonumber c_{{38}}&=&-\frac{1}{\sqrt {3}}
    {\cal D}_{{55}}+\frac{1}{\sqrt {3}}{\cal D} _{{77}}+\frac{1}{\sqrt {3}}
    {\cal D}_{{66}}-\frac{1}{\sqrt {3}}{\cal D}_{{44}} +\frac{2}{3}{\cal D}_{{38}}
\label{inverseclrelation}
\ea

The simplified form of the $c_{ij}$ matrix given in (\ref{cmatrix})
implies a matrix  $L_{ij}$ of the form:
 \ba
    L = \left(%
\begin{array}{cccccccc}
  {\cal D}_{11} & -\Delta_{12} & 0 & 0 & 0 & 0 & 0 & 0 \\
  \Delta_{12} & {\cal D}_{22} & 0 & 0 & 0 & 0 & 0 & 0 \\
  0 & 0 & {\cal D}_{33} & 0 & 0 & 0 & 0 & {\cal D}_{38} \\
  0 & 0 & 0 & {\cal D}_{44} & -\Delta_{13} & 0 & 0 & 0 \\
  0 & 0 & 0 & \Delta_{13} & {\cal D}_{55} & 0 & 0 & 0 \\
  0 & 0 & 0 & 0 & 0 & {\cal D}_{66} & -\Delta_{23} & 0 \\
  0 & 0 & 0 & 0 & 0 & \Delta_{23} & {\cal D}_{77} & 0 \\
  0 & 0 & {\cal D}_{83} & 0 & 0 & 0 & 0 & {\cal D}_{88} \\
\end{array}%
\right)
 \ea
where we have used the notation
$\Delta_{ij}=\frac{m_i^2-m_j^2}{2p}$.

The corresponding eigenvalues are:
  \ba
    \lambda_{1}&=&\frac{1}{2}[({\cal D}_{11}+{\cal D}_{22})-\sqrt{({\cal D}_{22}-{\cal D}_{11})^2
    -4\Delta_{12}^{2}}]
    \equiv \frac{1}{2}[({\cal D}_{11}+{\cal D}_{22})-\Omega_{12}]
   \nonumber  \\   \lambda_{2} &=& \frac{1}{2}[({\cal D}_{11}+{\cal D}_{22})+
     \sqrt{({\cal D}_{22}-{\cal D}_{11})^2-4\Delta_{12}^{2}}]
    \equiv \frac{1}{2}[({\cal D}_{11}+{\cal D}_{22})+\Omega_{12}]
  \nonumber   \\   \lambda_{3}&=&\frac{1}{2}[({\cal D}_{33}+{\cal D}_{88})-
     \sqrt{({\cal D}_{33}-{\cal D}_{88})^2+4{\cal D}_{38}^{2}}]
    \equiv \frac{1}{2}[({\cal D}_{11}+{\cal D}_{22})-\Omega_{38}]
 \nonumber   \\ \lambda_{4}&=&\frac{1}{2}[({\cal D}_{44}+{\cal D}_{55})-\sqrt{({\cal D}_{44}
      -{\cal D}_{55})^2-4\Delta_{13}^{2}}]
    \equiv \frac{1}{2}[({\cal D}_{44}+{\cal D}_{55})-\Omega_{13}]
 \nonumber   \\ \lambda_{5}&=&\frac{1}{2}[({\cal D}_{44}+{\cal D}_{55})+
      \sqrt{({\cal D}_{44}-{\cal D}_{55})^2-4\Delta_{13}^{2}}]
    \equiv \frac{1}{2}[({\cal D}_{44}+{\cal D}_{55})+\Omega_{13}]
\nonumber   \\  \lambda_{6}&=&\frac{1}{2}[({\cal D}_{66}+{\cal D}_{77})
   -\sqrt{({\cal D}_{66}-{\cal D}_{77})^2-4\Delta_{23}^{2}}]
    \equiv \frac{1}{2}[({\cal D}_{66}+{\cal D}_{77})-\Omega_{23}]
 \nonumber  \\ \lambda_{7}&=&\frac{1}{2}[({\cal D}_{66}+{\cal D}_{77})
   +\sqrt{({\cal D}_{66}-{\cal D}_{77})^2-4\Delta_{23}^{2}}]
    \equiv \frac{1}{2}[({\cal D}_{66}+{\cal D}_{77})+\Omega_{23}]
 \nonumber  \\  \lambda_{8}&=&\frac{1}{2}[({\cal D}_{33}+{\cal D}_{88})+
     \sqrt{({\cal D}_{33}-{\cal D}_{88})^2+4{\cal D}_{38}^{2}}]
    \equiv \frac{1}{2}[({\cal D}_{11}+{\cal D}_{22})+\Omega_{38}].
 \ea
The probability of a
neutrino of flavor $\nu_{\alpha}$, created at time $t=0$, being converted to a
flavor $\nu_{\beta}$ at a later time t, is calculated in the
Lindblad framework~\cite{lindblad,gago} to be
 \ba
    P_{\nu_{\alpha}\rightarrow \nu_{\beta}}(t)=\Tr [\rho^{\alpha}(t)\rho^{\beta}]
    = \frac{1}{3}+ \frac{1}{2}\sum_{i,j,k}
    e^{\lambda_{k}t}D_{ik}D_{kj}^{-1}
    \rho_{j}^{\alpha}(0)\rho_{i}^{\beta}~.
\label{lindbladprob}
 \ea
where the matrix \textbf{D}  and its inverse are \ba
    \textbf{D}=
\left(%
\begin{array}{cccccccc}
  \frac{\lambda_{1}-{\cal D}_{22}}{\Delta_{12}} & \frac{\lambda_{2}-{\cal D}_{22}}{\Delta_{12}} & 0 & 0 & 0 & 0 & 0 & 0 \\
  1 & 1 & 0 & 0 & 0 & 0 & 0 & 0 \\
  0 & 0 & \frac{\lambda_3-{\cal D}_{33}}{{\cal D}_{38}} & 0 & 0 & 0 & 0 & \frac{\lambda_8-{\cal D}_{33}}{{\cal D}_{38}} \\
  0 & 0 & 0 & \frac{\lambda_{4}-{\cal D}_{55}}{\Delta_{13}} & \frac{\lambda_{5}-{\cal D}_{55}}{\Delta_{13}} & 0 & 0 & 0 \\
  0 & 0 & 0 & 1 & 1 & 0 & 0 & 0 \\
  0 & 0 & 0 & 0 & 0 & \frac{\lambda_{6}-{\cal D}_{77}}{\Delta_{23}} & \frac{\lambda_{7}-{\cal D}_{77}}{\Delta_{23}} & 0 \\
  0 & 0 & 0 & 0 & 0 & 1 & 1 & 0 \\
  0 & 0 & 1 & 0 & 0 & 0 & 0 & 1 \\
\end{array}%
\right). \ea
and
\ba
\textbf{D}^{-1}=
\left(%
\begin{array}{cccccccc}
  -\frac{\Delta_{12}}{\Omega_{12}} & \frac{\lambda_{2}-{\cal D}_{22}}{\Omega_{12}} & 0 & 0 & 0 & 0 & 0 & 0 \\
  \frac{\Delta_{12}}{\Omega_{12}} & -\frac{\lambda_{1}-{\cal D}_{22}}{\Omega_{12}} & 0 & 0 & 0 & 0 & 0 & 0 \\
  0 & 0 & -\frac{{\cal D}_{38}}{\Omega_{38}} & 0 & 0 & 0 & 0 & \frac{\lambda_{8}-{\cal D}_{33}}{\Omega_{38}} \\
  0 & 0 & 0 & -\frac{\Delta_{13}}{\Omega_{13}} & \frac{\lambda_{5}-{\cal D}_{55}}{\Omega_{13}} & 0 & 0 & 0 \\
  0 & 0 & 0 & \frac{\Delta_{13}}{\Omega_{13}} & -\frac{\lambda_{4}-{\cal D}_{55}}{\Omega_{13}} & 0 & 0 & 0 \\
  0 & 0 & 0 & 0 & 0 & -\frac{\Delta_{23}}{\Omega_{23}} & \frac{\lambda_{7}-{\cal D}_{77}}{\Omega_{23}} & 0 \\
  0 & 0 & 0 & 0 & 0 & \frac{\Delta_{23}}{\Omega_{23}} & -\frac{\lambda_{6}-{\cal D}_{77}}{\Omega_{23}} & 0 \\
  0 & 0 & \frac{{\cal D}_{38}}{\Omega_{38}} & 0 & 0 & 0 & 0 & -\frac{\lambda_{3}-{\cal D}_{33}}{\Omega_{38}}  \\
\end{array}%
\right). \ea

It will be sufficient to
look explicitly at the $k=1$ and $k=2$ terms in the sum of the
right hand side of (\ref{lindbladprob}),
since by block symmetry the other terms will be of the same form.

For the $k=1$ term we have:
 \[
    e^{\lambda_1 t}\left(D_{i1}D^{-1}_{1j} \rho_{j}^{\alpha}(0)
    \rho_{i}^{\beta}\right)= e^{\lambda_1 t}\left(\rho_{1}^{\alpha}
    \rho_{1}^{\beta}
    D_{11}D^{-1}_{11} +\rho_{2}^{\alpha} \rho_{2}^{\beta}
    D_{21}D^{-1}_{12}+ \rho_{2}^{\alpha} \rho_{1}^{\beta}
    D_{11}D^{-1}_{12}+\rho_{1}^{\alpha} \rho_{2}^{\beta}
    D_{21}D^{-1}_{11}\right)
 \]

 \ba
    =e^{\frac{({\cal D}_{11}+{\cal D}_{22})t}{2}}
e^{\frac{-\Omega_{12}t}{2}}\left[ \rho_{1}^{\alpha}
    \rho_{1}^{\beta}
    \frac{-{\cal D}_{11}+{\cal D}_{22}+\Omega_{12}}{2\Omega_{12}}
    +\rho_{2}^{\alpha} \rho_{2}^{\beta} \frac{-{\cal D}_{22}+{\cal D}_{11}+\Omega_{12}}{2\Omega_{12}}\right.
    \\ \nonumber \left. + \rho_{2}^{\alpha} \rho_{1}^{\beta}
    \frac{\Delta_{12}}{\Omega_{12}}
    +\rho_{1}^{\alpha} \rho_{2}^{\beta}
    \frac{-\Delta_{12}}{\Omega_{12}}\right]\label{eqn1}
 \ea
Likewise for $k=2$ we have
\[
    e^{\lambda_2 t}\left(D_{i2}D^{-1}_{2j} \rho_{j}^{\alpha}(0)
    \rho_{i}^{\beta}\right)= e^{\lambda_2 t}\left(\rho_{1}^{\alpha} \rho_{1}^{\beta}
    D_{12}D^{-1}_{21} +\rho_{2}^{\alpha} \rho_{2}^{\beta}
    D_{22}D^{-1}_{22}+ \rho_{2}^{\alpha} \rho_{1}^{\beta}
    D_{12}D^{-1}_{22}+\rho_{1}^{\alpha} \rho_{2}^{\beta}
    D_{22}D^{-1}_{21}\right)
 \]
 \ba
    =e^{\frac{({\cal D}_{11}+{\cal D}_{22})t}{2}}e^{\frac{\Omega_{12}t}{2}}
    \left[\rho_{1}^{\alpha} \rho_{1}^{\beta}
    \frac{-{\cal D}_{22}+{\cal D}_{11}+\Omega_{12}}{2\Omega_{12}}
    +\rho_{2}^{\alpha} \rho_{2}^{\beta} \frac{(-{\cal D}_{22}+{\cal D}_{11}-\Omega_{12})}
     {2\Omega_{12}}\right.
    \\ \nonumber + \left.\rho_{2}^{\alpha} \rho_{1}^{\beta}
    \frac{-\Delta_{12}}{\Omega_{12}}
    +\rho_{1}^{\alpha} \rho_{2}^{\beta}
    \frac{\Delta_{12}}{\Omega_{12}}\right]\label{eqn2}
 \ea
Upon combining equations (\ref{eqn1}) and (\ref{eqn2}) we obtain:
 \ba
(\ref{eqn1}) + (\ref{eqn2})=e^{({\cal D}_{11}+{\cal D}_{22})\frac{t}{2}}
    \left[\left(\rho_{1}^{\alpha} \rho_{1}^{\beta}+ \rho_{2}^{\alpha} \rho_{2}^{\beta}\right)
    \left(\frac{e^{-\Omega_{12}\frac{t}{2}}+ e^{\Omega_{12}\frac{t}{2}}}{2}\right)\right. \qquad
     \qquad \qquad \qquad
    \\ \nonumber \left.+\left(\frac{2\Delta_{12}(\rho_{1}^{\alpha}
     \rho_{2}^{\beta} - \rho_{2}^{\alpha}
    \rho_{1}^{\beta}) + \Delta {\cal D}_{21}
    (\rho_{1}^{\alpha} \rho_{1}^{\beta} - \rho_{2}^{\alpha}
    \rho_{2}^{\beta})}{\Omega_{12}}\right)
    \left(\frac{e^{-\Omega_{12}\frac{t}{2}}- e^{\Omega_{12}\frac{t}{2}}}{2}\right) \right]
 \ea
 As mentioned earlier,
by block symmetry we can see that the other terms will be of the same
 form.

We thus obtain for the relevant probability:
 \ba
    \nonumber P_{\nu_{\alpha}\rightarrow \nu_{\beta}}(t)&=& \frac{1}{3}
    + \frac{1}{2}\left\{ \left[ \left(\rho_{1}^{\alpha}
    \rho_{1}^{\beta} + \rho_{2}^{\alpha} \rho_{2}^{\beta}\right)
    \left(\frac{e^{-\Omega_{12}\frac{t}{2}}+
    e^{\Omega_{12}\frac{t}{2}}}{2}\right)\right.\right.
    \\ \nonumber &+&\left. \left(\frac{2\Delta_{12}(\rho_{1}^{\alpha}
     \rho_{2}^{\beta} - \rho_{2}^{\alpha}
    \rho_{1}^{\beta}) + \Delta {\cal D}_{21}
    \left(\rho_{1}^{\alpha} \rho_{1}^{\beta} - \rho_{2}^{\alpha}
    \rho_{2}^{\beta}\right)}{\Omega_{12}}\right)
    \left(\frac{e^{-\Omega_{12}\frac{t}{2}}-
    e^{\Omega_{12}\frac{t}{2}}}{2}\right)\right]
    e^{({\cal D}_{11}+{\cal D}_{22})\frac{t}{2}}
    \\ \nonumber &+&\left[ \left(\rho_{4}^{\alpha} \rho_{4}^{\beta}+
     \rho_{5}^{\alpha}
    \rho_{5}^{\beta}\right)
    \left(\frac{e^{-\Omega_{13}\frac{t}{2}}+
    e^{\Omega_{13}\frac{t}{2}}}{2}\right) \right.
    \\ \nonumber &+&\left. \left(\frac{2\Delta_{13}\left(\rho_{4}^{\alpha}
     \rho_{5}^{\beta} - \rho_{5}^{\alpha}
    \rho_{4}^{\beta}\right) + \Delta {\cal D}_{54}
    \left(\rho_{4}^{\alpha} \rho_{4}^{\beta} - \rho_{5}^{\alpha}
    \rho_{5}^{\beta}\right)}{\Omega_{13}}\right)
    \left(\frac{e^{-\Omega_{13}\frac{t}{2}}- e^{\Omega_{13}
    \frac{t}{2}}}{2}\right) \right]
    e^{({\cal D}_{44}+{\cal D}_{55})\frac{t}{2}}
    \\ \nonumber &+& \left[ \left(\rho_{6}^{\alpha} \rho_{6}^{\beta}+
     \rho_{7}^{\alpha} \rho_{7}^{\beta}\right)
    \left(\frac{e^{-\Omega_{23}\frac{t}{2}}+
    e^{\Omega_{23}\frac{t}{2}}}{2}\right)\right.
    \\ \nonumber &+& \left. \left(\frac{2\Delta_{23}(\rho_{6}^{\alpha}
     \rho_{7}^{\beta} -
    \rho_{7}^{\alpha} \rho_{6}^{\beta}) + \Delta {\cal D}_{76}
    \left(\rho_{6}^{\alpha} \rho_{6}^{\beta} - \rho_{7}^{\alpha}
    \rho_{7}^{\beta}\right)}{\Omega_{23}}\right)
    \left(\frac{e^{-\Omega_{23}\frac{t}{2}}- e^{\Omega_{23}
    \frac{t}{2}}}{2}\right) \right]
    e^{({\cal D}_{66}+{\cal D}_{77})\frac{t}{2}}
    \\  \nonumber &+& \left[ \left(\rho_{3}^{\alpha} \rho_{3}^{\beta}+
     \rho_{8}^{\alpha} \rho_{8}^{\beta}\right)
    \left(\frac{e^{-\Omega_{38}\frac{t}{2}}+
    e^{\Omega_{38}\frac{t}{2}}}{2}\right)\right.
    \\ \nonumber &+& \left. \left(\frac{2{\cal D}_{38}(\rho_{3}^{\alpha}
     \rho_{8}^{\beta} -
    \rho_{8}^{\alpha} \rho_{3}^{\beta}) + \Delta {\cal D}_{83}
    \left(\rho_{3}^{\alpha} \rho_{3}^{\beta} - \rho_{8}^{\alpha}
    \rho_{8}^{\beta}\right)}{\Omega_{38}}\right)
    \left(\frac{e^{-\Omega_{38}\frac{t}{2}}- e^{\Omega_{38}
    \frac{t}{2}}}{2}\right) \right]
    e^{({\cal D}_{33}+{\cal D}_{88})\frac{t}{2}}.
 \ea
Above we have used the notation that $\Delta {\cal D}_{ij}={\cal D}_{ii}-{\cal D}_{jj}$.
We have assumed that
 $2|\Delta_{ij}|<|\Delta {\cal D}_{ij}|$ with the consequence that $\Omega_{ij}$,
$ij=12,13,23$ is imaginary. However,
$\Omega_{38}=\sqrt{({\cal D}_{33}-{\cal D}_{88})^2+4{\cal D}_{38}^{2}}$ will be
real.
Thus, the final expression for the probability reads
 \ba
    \nonumber P_{\nu_{\alpha}\rightarrow \nu_{\beta}}(t)&=& \frac{1}{3}
    + \frac{1}{2}\left\{ \left[ \rho_{1}^{\alpha}
    \rho_{1}^{\beta}
    \cos\left(\frac{|\Omega_{12}|t}{2}\right)
     + \left(\frac{ \Delta {\cal D}_{21}
   \rho_{1}^{\alpha} \rho_{1}^{\beta} }{|\Omega_{12}|}\right)
    \sin\left(\frac{|\Omega_{12}|t}{2}\right)\right]
    e^{({\cal D}_{11}+{\cal D}_{22})\frac{t}{2}}\right.
    \\ \nonumber &+&\left[ \rho_{4}^{\alpha} \rho_{4}^{\beta}
    \cos\left(\frac{|\Omega_{13}|t}{2}\right) + \left(\frac{\Delta {\cal D}_{54}
    \rho_{4}^{\alpha} \rho_{4}^{\beta} }{|\Omega_{13}|}\right)
    \sin\left(\frac{|\Omega_{13}|t}{2}\right) \right]
    e^{({\cal D}_{44}+{\cal D}_{55})\frac{t}{2}}
    \\ \nonumber &+& \left[ \rho_{6}^{\alpha} \rho_{6}^{\beta}
    \cos\left(\frac{|\Omega_{23}|t}{2}\right)
     +\left(\frac{\Delta {\cal D}_{76}
    \rho_{6}^{\alpha} \rho_{6}^{\beta} }{|\Omega_{23}|}\right)
    \sin\left(\frac{|\Omega_{23}|t}{2}\right) \right]
    e^{({\cal D}_{66}+{\cal D}_{77})\frac{t}{2}}
    \\ \nonumber  &+& \left[ \left(\rho_{3}^{\alpha} \rho_{3}^{\beta}+
     \rho_{8}^{\alpha} \rho_{8}^{\beta}\right)
    \cosh\left(\frac{\Omega_{38}t}{2}\right)\right.
    \\ \nonumber &+& \left.  \left(\frac{2{\cal D}_{38}(\rho_{3}^{\alpha}
     \rho_{8}^{\beta} -
    \rho_{8}^{\alpha} \rho_{3}^{\beta}) + \Delta {\cal D}_{83}
    \left(\rho_{3}^{\alpha} \rho_{3}^{\beta} - \rho_{8}^{\alpha}
    \rho_{8}^{\beta}\right)}{\Omega_{38}}\right)
    \sinh\left(\frac{\Omega_{38}t}{2}\right)\right]
    e^{({\cal D}_{33}+{\cal D}_{88})\frac{t}{2}}~. \\
\label{finalnuprob}
 \ea
On using the relations
 \ban
    \rho^{\alpha}_{0} &=& \sqrt{\frac{2}{3}}
    \\ \rho^{\alpha}_{1} &=& 2 Re(U_{\alpha 1}^{*}U_{\alpha 2})
    \\ \rho^{\alpha}_{2} &=& -2 Im(U_{\alpha 1}^{*}U_{\alpha 2})
    \\ \rho^{\alpha}_{3} &=& |U_{\alpha 1}|^{2}-|U_{\alpha 2}|^{2}
    \\ \rho^{\alpha}_{4} &=& 2 Re(U_{\alpha 1}^{*}U_{\alpha 3})
    \\ \rho^{\alpha}_{5} &=& -2 Im(U_{\alpha 1}^{*}U_{\alpha 3})
    \\ \rho^{\alpha}_{6} &=& 2 Re(U_{\alpha 2}^{*}U_{\alpha 3})
    \\ \rho^{\alpha}_{7} &=& -2 Im(U_{\alpha 2}^{*}U_{\alpha 3})
    \\ \rho^{\alpha}_{8} &=&\sqrt{\frac{1}{3}} \left( |U_{\alpha 1}|^{2}
    +|U_{\alpha 2}|^{2}- 2|U_{\alpha 3}|^{2}\right).
 \ean
we can readily see that for a real $U$ matrix (i.e. no CP violating phases) the relevant probabilities are bounded.
Indeed, there is no
danger of the $\cosh$ and $\sinh$ terms blowing up with time
$t$, as we always have
$\Omega_{38}<{\cal D}_{33}+{\cal D}_{88}$. We can see this by checking that
$({\cal D}_{33}-{\cal D}_{88})^2+4{\cal D}_{38}^{2}<{\cal D}_{33}^2+{\cal D}_{88}^2+2{\cal D}_{33}{\cal D}_{88}~$,
 $-2{\cal D}_{33}{\cal D}_{88}+4{\cal D}_{38}^2<2{\cal D}_{33}{\cal D}_{88}~,$
 and ${\cal D}_{38}^2<{\cal D}_{33}{\cal D}_{88}$,
 which are automatically satisfied, as can be
readily checked from the relevant expressions (\ref{clrelation}).

We are now ready to discuss the fit to the experimental data.
This is done in the next section.

\section{Fitting the Experimental Data}

In order to check the viability of our simplified scenario, we have performed a
$\chi^2 $ comparison (as opposed to a $\chi^2 $ fit)
to SuperKamiokande sub-GeV and multi GeV data
(the 40 data points that are shown in Figure 1), CHOOZ
data (15 data points), KamLAND (13 data points, shown in Figure 2)
and LSND (1 datum),
for a sample point in the
vast parameter space of our
extremely simplified version of decoherence models.
Rather than performing a $\chi^2$-fit (understood as a run over all the
parameter space to find the global minimum of the $\chi^2 $ function) we have selected  (by ``eye'' and not by $\chi$) a point which is not optimised to give
the best fit to the existing data. Instead, our sample point  must be regarded
as a local minimum around a starting point chosen by an educated guess.
It follows then that  it may be quite  possible to find a better fitting point through a
complete (and highly time consuming) scan over the whole parameter
space.

To simplify the analysis and gain intuition concerning the rather cumbersome expressions for the transition probabilities,
we have imposed,
\ba
{\cal D}_{11} =  {\cal D}_{22}~,~\qquad &&  {\cal D}_{44}={\cal D}_{55}~, \nonumber \\
{\cal D}_{66}  = {\cal D}_{77}~,~\qquad && {\cal D}_{33}={\cal D}_{88}~, \nonumber \\
  {\cal D}_{38}={\cal D}_{83}&=&0~.
\label{general}
\ea
implying a diagonal ${\cal D}$-matrix.

Later on we shall
set some of the ${\cal D}_{ii}$ to zero. Furthermore, we have also set the CP violating phase of the KMS matrix to zero, so that all the mixing matrix elements become real.

With these assumptions, the complicated
expression for the transition
probability (\ref{finalnuprob})
simplifies to:
\ba
    \nonumber P_{\nu_{\alpha}\rightarrow \nu_{\beta}}(t)&=& \frac{1}{3}
    + \frac{1}{2}\left\{  \rho_{1}^{\alpha}
    \rho_{1}^{\beta}
    \cos\left(\frac{|\Omega_{12}|t}{2}\right)
    e^{{\cal D}_{22}t } \right.
    \\ \nonumber &+&  \rho_{4}^{\alpha} \rho_{4}^{\beta}
    \cos\left(\frac{|\Omega_{13}|t}{2}\right)
    e^{{\cal D}_{55} t }
   +  \rho_{6}^{\alpha} \rho_{6}^{\beta}
    \cos\left(\frac{|\Omega_{23}|t}{2}\right)
    e^{{\cal D}_{66} t }
    \\   &+& \left. \left(\rho_{3}^{\alpha} \rho_{3}^{\beta}+
     \rho_{8}^{\alpha} \rho_{8}^{\beta}\right)
    \cosh\left(\frac{\Omega_{38}t}{2}\right)
    e^{{\cal D}_{33} t} \right\}
 \ea
for both, neutrino and antineutrino sectors.

As indicated by the state-of-the-art analysis, masses and mixing angles are selected to have the values

\centerline{$\Delta m_{12}^2 =
7 \cdot 10^{-5}$~eV$^2$,}

\centerline{$\Delta m_{23}^2 =
2.5 \cdot 10^{-3}$~eV$^2$,}

\centerline{$\theta_{23}  = \pi/4,\;\;\;$ $\theta_{12} = .45,\;\;\;$$\theta_{13} = .05$}.

\noindent

For the decoherence parameters we find
\ba
{\cal D}_{33}={\cal D}_{66}=0~,~\qquad
{\cal D}_{11}={\cal D}_{22}={\cal D}_{44}={\cal D}_{55}=-\frac{\;\;\; 1.3 \cdot 10^{-2}\;\;\;}
{L},
\label{special}
\ea
in units of 1/km with $L=t$ the oscillation length. The
$1/L$-behaviour of ${\cal D}_{11} $, implies
oscillation-length independent Lindblad exponents. We shall attempt
to interpret this result in the next section.

The complete positivity of the case defined by
(\ref{general}) and (\ref{special}) is guaranteed; this follows from the
fact that the solutions for $c_{ij}$ in terms of ${\cal D}_{\ell k}$
(cf. (\ref{inverseclrelation})) is such that the only non-zero
elements are $c_{88}= c_{38}/\sqrt{3}$. Such a $C$-matrix has only non-negative eigenvalues
 \ba C-{\rm matrix~eigenvalues} =
\left(0,0,0,0,0,0,0,- 8 \;{\cal D}_{11} /3 \right)
\label{ceigen} \ea

In summary we have introduced only
one new parameter, a new degree of freedom,
by means of which we shall try
to explain all the available experimental data.
It is important to stress that the inclusion of one
new degree of freedom by itself
does not guarantee that all the experimental observations can be accounted for.
Indeed for situations without decoherence the addition of a sterile neutrino (which introduces
four new degrees of freedom -excluding CP violating phases)  seemed to be insufficient for
matching  {\it all}
available experimental data, at least in CPT conserving
situations~\cite{mohapatra}.

In order to test our model
with this decoherence parameter,
we have calculated the zenith angle dependence of the
ratio ``observed-events/(expected-events in the no oscillation case)'', for muon and
electron atmospheric neutrinos, for the sub-GeV and multi-GeV energy ranges,
when mixing is taken into account. Since matter effects are important for
atmospheric neutrinos, we have implemented them through a
two-shell model, where
the density in the mantle (core) is taken to be roughly 3.35 (8.44) gr/cm$^3$,
and the core radius is taken to be 2887 km.

The results are shown in Fig. 1,
where, for the sake of comparison,
we have also included
the experimental data.  As can be easily seen the agreement is remarkable.
\begin{figure}[tb]
\includegraphics[width=9cm]{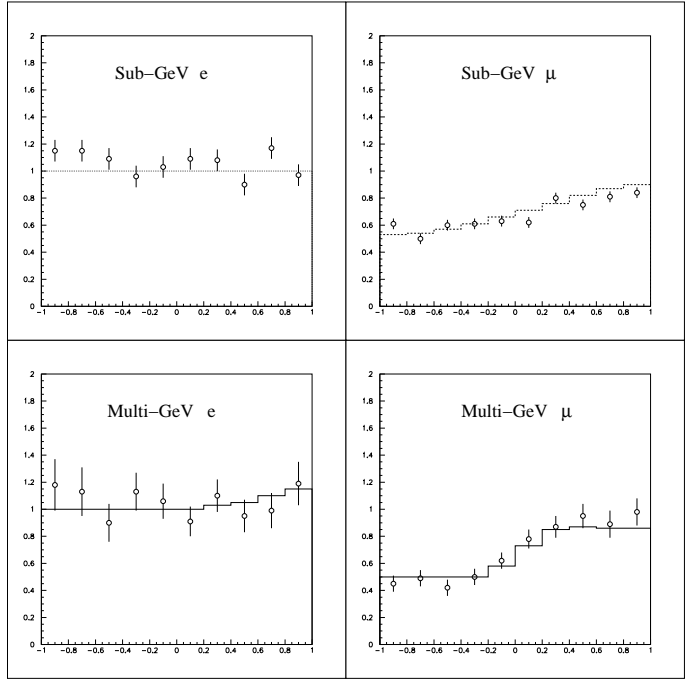} \hfill
\caption{Decoherence fit. The dots correspond to SK data.}
\label{bestfit}
\end{figure}

As bare eye comparisons can be misleading, we have also calculated the
$\chi^2$  value for each of the cases, defining the atmospheric
$\chi^2$  as
\begin{equation}
  \chi^2_{\rm atm}= \sum_{M, S}\sum_{\alpha=e,\mu}\sum_{i=1}^{10}
     \frac{(R_{\alpha,i}^{\rm exp}-
     R_{\alpha,i}^{\rm th})^2}{\sigma_{\alpha i}^2} \quad .
\end{equation}
Here $\sigma_{\alpha,i}$ are the statistical errors,
the ratios $R_{\alpha,i}$ between the observed and predicted signal
can be written as
\begin{equation}
  R_{\alpha,i}^{\rm exp}= N_{\alpha,i}^{\rm exp}/N_{\alpha,i}^{\rm MC}
\end{equation}
(with $\alpha$ indicating the lepton flavor and $i$ counting the
different bins, ten in total)
and $M,S$ stand for the multi-GeV and sub-GeV data respectively.
For the CHOOZ experiment we used the 15 data points with their
statistical errors, where in each bin we averaged the probability
over energy. For the KamLAND experiment, their 13 data points have
been used  for a fixed distance of $L_0= 180$ km, as if all antineutrinos
detected in KamLAND were due to a single reactor at this distance
and plotted in figure 2
while for  LSND one datum has been included.

\begin{figure}[tb]
\includegraphics[width=15cm]{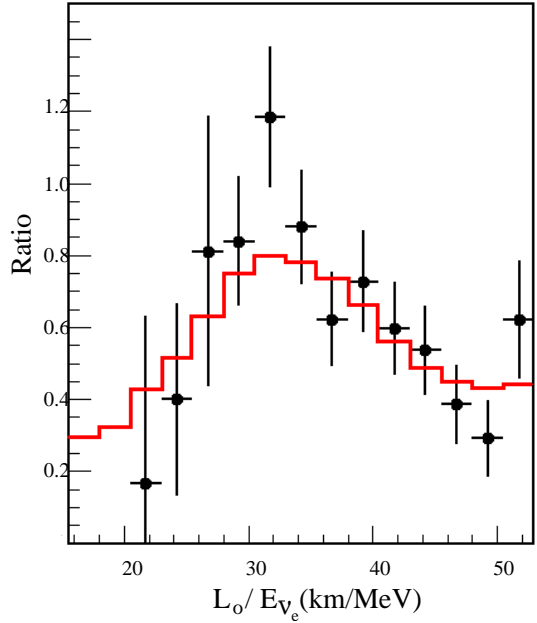} \hspace{-7cm}
\caption{Ratio of the observed $\overline{\nu}_e$ spectrum to the expectation
versus $L_0/E$ for our decoherence model. The dots correspond to KamLAND data}
\end{figure}

The results are summarised in Table 1, where we
present the $\chi^2$ comparison for the model in question
and the standard scenario (calculated with the same program).

\begin{table}[h]
\centering
\begin{tabular}{|c|c|c|}
\hline\hline
$\chi^2$ &  decoherence  & standard scenario  \\ [0.5 ex]
\hline\hline
SK  sub-GeV& 38.0  & 38.2 \\ \hline
SK Multi-GeV & 11.7  & 11.2 \\ \hline
Chooz & 4.5 & 4.5  \\\hline
 KamLAND & 16.7 & 16.6 \\\hline
LSND & 0.  & 6.8  \\\hline
TOTAL & 70.9  & 77.3 \\[1ex]
\hline\hline
\end{tabular}
\caption{$\chi^2$ obtained for our model and the one obtained in the standard
scenario for the different experiments calculated with the same program.}
\end{table}

{}From the table it becomes clear that our simplified version of decoherence in both
neutrino and antineutrino sectors can easily account for all the available
experimental information,
including LSND. It is important to stress once more that our sample
point was not obtained through a scan over all the parameter space,
but  by an educated guess, and therefore plenty of room is left
for improvements. On the other hand, for the mixing-only/no-decoherence
scenario,
we have taken the best fit values of the state of the art analysis
and therefore no significant improvements are expected.

As we have seen, the decoherence effects suffered by our model, are just
 an overall suppression  on some of the oscillatory terms modifying the
transition/survival probabilities at the per mil level (to account precisely for LSND,
a per mil evidence)
and therefore, no effect is expected (or found) in the oscillation dominated
physics, where the level of precision is at the percent level, at most.
We are guaranteed then to have an excellent agreement with solar data, as long
as we keep the relevant mass difference and mixing angle within
the LMA-I region, something which we certainly did. Thus, there is no need to
include these data on our fit.

At this point, a word of warning is in order. Although from the table,
it seems that the decoherence model we are presenting here and the
standard no-decoherence scenario provide equally good a fit, i.e.
while the former has a $ \chi^2 $/{\bf{DOF}} = 70.9/63 the latter
has a $\chi^2$/{\bf{DOF}} = 77.3/64, both  quite ``acceptable''
from the statistical point of view, one must remember that only the
decoherence model can explain the LSND result. This fact, however, gets
blurred in the total $\chi^2$ because LSND is represented by only one
experimental point with a poor precision.

Before closing this section, it is worth revisiting
the models of I, in order to understand
in the above context, the failure of complete
positivity in certain regions of the parameter space.
In that case, the following restrictions on
the decoherence matrix (which was also diagonal, as in the case
(\ref{general}) above) had been imposed~\cite{barenboim2}:
\ba
{\cal D}_{11} &= & {\cal D}_{22} = {\cal D}_{44}={\cal D}_{55} = -2\cdot 10^{-18} \cdot E =-A~,\nonumber \\
{\cal D}_{66} & =& {\cal D}_{77}= {\cal D}_{33}={\cal D}_{88} =
-10^{-24}/E=-B~,
\nonumber \\
{\cal D}_{38} &=& {\cal D}_{83}=0~, \ea leading to a solution for
the c-matrix   $c_{{38}}=\frac{2}{3}\,\sqrt {3}{\it
A}-\frac{2}{3}\,\sqrt {3}{\it B}, \; c_{{5 5}}=\frac{2}{3}\,{\it
B},\; c_{{44}}=\frac{2}{3}\,{\it B}, \;c_{{88}}=\frac{2}{3}\,{\it
A},\;c_{{ 66}}=\frac{2}{3}\,{\it B},\;c_{{22}}=\frac{2}{3}\,{\it
B},\;c_{{33}}=-4/3\,{\it B}+2 \,{\it A},\;c_{{77}}=\frac{2}{3}\,{\it
B},\;c_{{11}}=\frac{2}{3}\,{\it B}$ (all other matrix entries zero)
such that the pertinent  eigenvalues
$(-2\,{\it B}+\frac{8}{3}\,{\it A},\frac{2}{3}\,{\it
B},\frac{2}{3}\,{\it B},\frac{2}{3}\,{\it B},\frac{2}{3}\,{\it
B},\frac{2}{3}\,{\it B},\frac{2}{3}\,{\it B},\frac{2}{3}\,{\it B})$
which are not positive for arbitrary values of $A$ and $B$.
The positivity condition can be obtained
by demanding positivity of the first eigenvalue, i.e. $2 \cdot
10^{-24}/E<\frac{16}{3}\cdot 10^{-18} \cdot E$, where E is in units
of GeV, which leads to the condition $E>{\cal O} (1~{\rm MeV})$
which was  the condition found in I.

\section{Attempt at Interpreting the Fit}

The microscopic origin of the ``observed'' decoherence 
effects, according to our fit above ((\ref{special}),(\ref{ceigen})),  
may not be unique.
In fact there can be many contributions to the decoherence-induced
(oscillation-independent) damping (\ref{ceigen}), 
which modulates the oscillatory terms, arising from a variety 
of effects, ranging from microscopic quantum-space time fluctuations 
(`stochastic quantum gravity foam'~\cite{wheeler}), to ordinary
matter effects, e.g. uncertainties in the energy and oscillation 
length of the (anti)neutrino beam. It is the purpose of this
section to attempt and disentangle these very different 
in nature contributions, and in particular to estimate their 
plausible order in terms of microscopic theoretical models and see 
which one gives the order specified by the fit.
 
To understand the results (\ref{special}),(\ref{ceigen}), in
connection with either stochastically fluctuating quantum-gravity
space-time-foam models~\cite{poland,msw2}, 
or energy-uncertainty driven decoherence~\cite{ohlsson},
it suffices to restrict our discussion to
the simplified (but phenomenologically realistic ) case of three
neutrino families, but with dominant mixing only between 12,
23~\cite{msw2}, with mixing angles $\theta_{12}=\theta_{23} =
\theta$, and $\theta_{13} = 0$. The three generation case with full
mixing does not affect qualitatively the form of the damping
exponents used to fit of the oscillation experiments,
and hence we are free to use results on
theoretical models from this simpler case, in order to interpret
qualitatively the above result.

\subsection{Stochastic Quantum-Gravity Models}

We commence our analysis with models of 
neutrinos propagating in a quantum-gravity ground state. 
To this end, consider the propagation of such a neutrino system,
in a medium whose density stochastically fluctuates.
For our purposes the medium is taken to be a quantum space time 
foam~\cite{wheeler},
with fluctuating densities of
charged black-hole/anti-black-hole pairs produced by the vacuum and
being absorbed by it within Planckian time scales~\cite{barenboim}.
Such a case, will not produce any vacuum charge on average,
but the associated density fluctuations will produce vacuum fluctuations
in electron currents with which electron neutrinos will interact.
An inherent CPT violation may result in asymmetries between
particle antiparticle sectors, as far as the appropriate interactions
of the (anti)neutrinos with these currents are concerned.

Such asymmetries can produce a bias in flavour of, say, the electron
current fluctuations in the space-time foam vacuum, with the result
that a ``gravitationally-induced'' MSW~\cite{msw} effect is in
place~\cite{barenboim}, with small contributions to the standard
oscillations due to bare mass differences between neutrinos, that
could be due to non-gravitational physics~\footnote{In refs.
\cite{barenboim,msw2} we go one step further and consider cases
where the gravitational foam is responsible for the entire portion
of the observed mass differences, and not only a small part as done
here. However, for our purposes here this point is not relevant.}.
The precise reason why a preference to electron current fluctuations
as opposed to positron ones cannot be answered at this stage, since
a microscopic model of quantum space time foam is lacking. However,
the situation is not incompatible with the intrinsic CPT violation
(to be precise {\it microscopic time irreversibility}, unrelated to
CP properties, according to which the generator of time reversal
operations is ill-defined) characterising such problems, where a
proper scattering matrix cannot be defined~\cite{wald}.

In such a case, the evolution equation of the density matrix
$\rho$ of the neutrino
probe involves a time-reversal (CPT) breaking
decoherence matrix of a double commutator form~\cite{loreti},
\ba
\partial_t \rho = i[\rho, H_{eff}] - \Omega^2 [H_I, [H_I, \rho]]
\label{double} \ea where $\langle n(r) n(r') \rangle = \Omega^2n_0^2
\delta (r - r') $ denote the stochastic (Gaussian) fluctuations of
the density of the medium and $H_{eff}=H_0+H_I$, $H_0$ being the
standard Hamiltonian, and $H_I$ an MSW-type interaction~\cite{msw,loreti}.
This double-commutator decoherence is a
specific case of Lindblad evolution, of the type considered in previous
sections, with a $C$-matrix of the form:

\begin{equation}
C=\left(%
\begin{array}{cccccccc}
  h_1^2 & 0 & h_1h_3 & 0 & 0 & 0 & 0 & h_1h_8 \\
  0 & 0 & 0 & 0 & 0 & 0 & 0 & 0 \\
  h_1h_3 & 0 & h_3^2 & 0 & 0 & 0 & 0 & h_3h_8 \\
  0 & 0 & 0 & 0 & 0 & 0 & 0 & 0 \\
  0 & 0 & 0 & 0 & 0 & 0 & 0 & 0 \\
  0 & 0 & 0 & 0 & 0 & 0 & 0 & 0 \\
  0 & 0 & 0 & 0 & 0 & 0 & 0 & 0 \\
  h_1h_8 & 0 & h_3h_8 & 0 & 0 & 0 & 0 & h_8^2 \\
\end{array}%
\right)
\end{equation}
with $ h_1=(a_{\nu_e}-a_{\nu_{\mu}})\sin(2\theta),\;
h_3=(a_{\nu_e}-a_{\nu_{\mu}})\cos(2\theta),$ and
$h_8=\frac{(a_{\nu_e}-a_{\nu_{\mu}})}{\sqrt{3}}~.$ This matrix
indeed has positive eigenvalues for real $h_i$. The difference
$a_{\nu_e}-a_{\nu_\mu}$ is proportional to the average density of
the medium $n_0$.

We note at this stage that,
for gravitationally-induced MSW effects (due to, say, black-hole foam
models as in \cite{barenboim}), one may write
\ba
\Delta a_{e\mu} \equiv a_{\nu_e}-a_{\nu_\mu} \propto G_N n_0
\ea
with $G_N=1/M_P^2$, $M_P \sim 10^{19}~{\rm GeV}$, the four-dimensional
Planck scale. This gravitational coupling replaces the weak interaction
Fermi coupling constant $G_F$ in the conventional MSW effect.
This is the case we shall be interested in for the purposes of this work.

In such a  case the density fluctuations $\Omega^2$ are therefore assumed
small compared to other quantities present in the formulae, and an expansion to leading order in $\Omega^2$ is in place.
Following then the standard
analysis, outlined above, one obtains the following expression for
the neutrino transition probability $\nu_e \leftrightarrow
\nu_\mu$ in this case, to leading order in the small
parameter $\Omega^2 \ll 1$:
\begin{eqnarray}
    P_{\nu_e\to \nu_{\mu}}&&=   \nonumber \\
    && e^{-\Delta a_{e\mu}^2\Omega^2t(1+\frac{\Delta_{12}^2}{4\Gamma}
(\cos(4\theta)-1))}
    \sin(t\sqrt{\Gamma})\cos^2(\theta)\sin^2(2\theta)\Delta
a_{e\mu}^2\Omega^2\Delta_{12}^2
    \left(\frac{3\sin^2(2\theta)\Delta_{12}^2}{4\Gamma^{5/2}}
-\frac{1}{\Gamma^{3/2}}\right) \nonumber \\
    && -e^{-\Delta
    a_{e\mu}^2\Omega^2t(1+\frac{\Delta_{12}^2}{4\Gamma}(\cos(4\theta)-1))}
    \cos(t\sqrt{\Gamma})
\cos^2(\theta)\sin^2(2\theta)\frac{\Delta_{12}^2}{2\Gamma}  \nonumber \\
    &&-e^{-\frac{\Delta a_{e\mu}^2\Omega^2\Delta_{12}^2t\sin^2(2\theta)}{\Gamma}}
    \cos^2(\theta)\frac{(\Delta a_{e\mu}+\cos(2\theta)\Delta_{12})^2}{2\Gamma}+\frac{1}{2}\cos^2(\theta)
\label{3genprob}
\end{eqnarray}
 where $\Gamma= (\Delta a_{e\mu}\cos(2\theta)+\Delta_{12})^2+\Delta
a_{e\mu}^2\sin^2(2\theta)~,$ $\Delta_{12}=\frac{\Delta m_{12}^2}{2p}~.$

{}From (\ref{3genprob}) we easily conclude that the exponents of the
damping factors due to the stochastic-medium-induced decoherence,
are therefore of the generic form, for $t = L$, the oscillation
length (in units of $c=1$): \ba {\rm exponent} \sim \Delta
a_{e\mu}^2\Omega^2t\left(1+\frac{\Delta_{12}^2(\cos(4\theta)-1)}{4\Gamma}\right)
\label{gammadelta} \ea The reader should note at this stage that, in
the limit $\Delta_{12}\to 0$, which could characterise the situation
in \cite{barenboim2}, where the space-time foam effects on the
induced neutrino mass difference are the dominant ones, the damping
factor is of the form $ {\rm exponent}_{{\rm gravitational~MSW}}
\sim \Omega^2 (\Delta a_{e\mu})^2 L~,$ with the precise value of the
mixing angle $\theta$ not affecting the leading order of the various
exponents. However, in that case, as follows from (\ref{3genprob}),
the overall oscillation probability is suppressed by factors
proportional to $\Delta_{12}^2 $, and, hence, the stochastic
gravitational MSW effect~\cite{barenboim}, although in principle
capable of inducing mass differences for neutrinos, however does not
suffice to produce the bulk of the oscillation probability, which is
thus attributed to conventional flavour physics.

In what follows, therefore, we assume the case where $\Delta_{12}
\gg \Delta a_{e\mu}$, and this is the case we shall compare with the
results of our experimental fit above. The result of the fit
(\ref{special}),(\ref{ceigen}), then, implies that the above
decoherence-induced damping exponent (\ref{gammadelta}) is
independent of $L$ and actually we have, to leading order in $\Delta
a_{e\mu}/\Delta_{12} \ll 1$ (re-instating dimensions of $\hbar, c$):
\ba \Omega^2 (\Delta a_{e\mu})^2 \left(1 + \frac{{\rm
cos}(4\theta)-1}{4}\right) \cdot  L \sim  2.56\times 10^{-19}~{\rm
GeV}\cdot {\rm km}. \label{limitfit} \ea This in turn implies that
in this specific model of foam, the density fluctuations of the
space-time charged black holes is such that, for maximal mixing,
say, $\theta = \pi/4$ assumed for concreteness, and for $L \sim
180$~Km, as appropriate for the KamLAND experiment, the decoherence
damping factor 
is $ {\cal D} = \Omega^2 G_N^2 n_0^2 \sim 2.84 \cdot
10^{-21}~{\rm GeV}$, if the result of the fit is due exclusively to
this effect (note that the mixing angle part does not affect the
order of the exponent). Smaller values are found for longer $L$,
such as in atmospheric neutrino experiments. 

The independence of the
relevant damping exponent from the oscillation length, then, implied
by our fit above, may be understood as follows in this context: 
In the spirit of \cite{barenboim}, 
the quantity $G_N n_0 = \xi \frac{\Delta m^2}{E}$,
where $\xi \ll 1$ parametrises the contributions of the foam to the
induced neutrino mass differences, according to our discussion
above. Hence, the damping exponent becomes in this case $ \xi^2
\Omega^2 (\Delta m^2)^2 \cdot L /E^2 $. Thus, for oscillation
lengths $L$ we have 
$L^{-1} \sim \Delta m^2/E$, and one is left with  the following 
estimate for the dimensionless quantity $\xi^2
\Delta m^2 \Omega^2/E \sim 1.3 \cdot 10^{-2}$. This  
implies that the quantity $\Omega^2$ is proportional to the
probe energy $E$. In principle, 
this is not an unreasonable result, and it is in
the spirit of \cite{barenboim}, since back reaction effects onto
space time, which affect the stochastic fluctuations $\Omega^2$, are
expected to increase with the probe energy $E$. However, 
due to the smallness of the quantity $\Delta m^2/E$, for energies 
of the order of GeV, and $\Delta m^2 \sim 10^{-3}$ eV$^2$, 
we conclude (taking into account that 
$\xi \ll 1$) that $\Omega^2$ in this case 
would be unrealistically large
for a quantum-gravity effect in the model. 

We remark at this point that, in such a model,
we can in principle bound independently the $\Omega$
and $n_0$ parameters by looking at the modifications induced by the
medium in the arguments of the oscillatory functions of the
probability (\ref{3genprob}), that is the period of oscillation.
Unfortunately this is too small to be detected in the above example,
for which $h_i \ll \Delta_{12}$.

The result of the fit, however, may be interpreted more generally, as
implying independent of the oscillation length $L$ exponents in the
decoherence exponential suppression factors in front of the
oscillatory terms in the transition probabilities.
In this sense, the bound (\ref{special}),(\ref{ceigen}) determined by
the fit above, can be applied to other stochastic decoherence models,
for instance the one discussed in \cite{msw2}, in which one averages
over random (Gaussian) fluctuations of the background space-time metric
over which the neutrino propagates.

In such an approach, one considers merely the Hamiltonian of the neutrino
in a stochastic metric background. This is one contribution to
decoherence, since other possible non-Hamiltonian
(like the Lindblad terms above)
interactions of the neutrino with the foam
are ignored.
In this case, one obtains transition probabilities
with exponential damping factors in front of the oscillatory
terms, but now the scaling with the oscillation length (time) is
quadratic~\cite{msw2}, consistent with time reversal invariance of the
neutrino Hamiltonian.  For instance, for the two generation case, which
suffices for our qualitative purposes in this work, we have:
\begin{eqnarray}
&&\langle e^{i(\omega_1-\omega_2)t}\rangle= e^{i\frac{{\left( {z_0^ +   - z_0^ -  } \right)t}}{k}}
e^{-\frac{1}{2}\left(-i\sigma_1 t\left(\frac{(m_1^2-m_2^2)}{k}+
V\cos2\theta\right)\right)} \times \nonumber \\
&&  e^{-\frac{1}{2}\left(\frac{i\sigma_2t}{2}\left( \frac{(m_1^2-m_2^2)}{k}+V\cos2\theta \right)
    -\frac{i\sigma_3t}{2}V\cos2\theta\right)} \times
    \nonumber \\&& e^{-\left(\frac{(m_1^2-m^2_2)^2}{2k^2}
   (9\sigma_1+\sigma_2+\sigma_3+\sigma_4)+\frac{2V\cos2\theta(m_1^2-m_2^2)}{k}
 (12\sigma_1+2\sigma_2-2\sigma_3)\right)t^2} \label{gravstoch}
 \end{eqnarray}
where $k$ is the neutrino energy, $\sigma_i~, i=1, \dots 4$
parameterise appropriately the stochastic fluctuations of the metric
in the model of \cite{msw2}, $\Upsilon  = \frac{{Vk}}{{m_1^2  -
m_2^2 }}$, $\left| \Upsilon  \right| \ll 1$, and $k^2  \gg m_1^2~,~m_2^2 $, and
 \begin{eqnarray}
 z_0^ +   &=& \frac{1}{2}\left(m_1^2  + \Upsilon (1 + \cos 2\theta )(m_1^2  - m_2^2 ) + \Upsilon ^2 (m_1^2  - m_2^2 )\sin ^2 2\theta \right) \nonumber \\
 z_0^ -   &=& \frac{1}{2}\left(m_2^2  + \Upsilon (1 - \cos 2\theta )(m_1^2  - m_2^2 ) - \Upsilon ^2 (m_1^2  - m_2^2 )\sin ^2 2\theta \right)~.
 \end{eqnarray}
Note that the metric fluctuations-$\sigma_i$ induced modifications
of the oscillation period, as well as exponential $e^{-(...)t^2}$
time-reversal invariant damping factors~\cite{msw2}. The latter is
attributed to the fact that in this approach, only the Hamiltonian
terms are taken into account (in a stochastically fluctuating metric
background), and as such time reversal invariance $t \to -t$ is not
broken explicitly. But there is of course decoherence, and the
associated damping.

We, then, observe that the
result of the fit above, (\ref{special}),(\ref{ceigen}),
implying $L$-independent exponents in the associated
damping factors due to decoherence,
may also apply to this case, implying for the damping
exponent:
\ba
\left(\frac{(m_1^2-m^2_2)^2}{2k^2}
   (9\sigma_1+\sigma_2+\sigma_3+\sigma_4)+
\frac{2V\cos2\theta(m_1^2-m_2^2)}{k}
 (12\sigma_1+2\sigma_2-2\sigma_3)
\right)t^2 \sim 1.3  \cdot 10^{-2}~. \nonumber \\
\ea Ignoring subleading MSW effects $V$, for simplicity,
and considering oscillation lengths $t=L \sim
\frac{2k}{(m_1^2-m^2_2)}$, we then observe that the independence of
the length $L$ result of the experimental fit, found above, may be
interpreted, in this case, as bounding the stochastic fluctuations
of the metric to $9\sigma_1+\sigma_2+\sigma_3+\sigma_4 \sim
1.3. \cdot 10^{-2}$. This is too large to be a quantum gravity
effect, which means that the $L^2$ contributions to the
damping due to stochastic fluctuations of the metric,
as in the model of \cite{msw2} above, cannot be the explanation of the fit.

\subsection{Conventional Explanation: Energy Uncertainties}

The reader's attention is called at this point to the fact that
such time-reversal invariance decoherence may also be due to
ordinary uncertainties~\cite{ohlsson} in energies and/or oscillation
lengths, which are unrelated to quantum gravity effects.
For instance, consider the ordinary oscillation formula for neutrinos,
with a mixing matrix $U$,
\begin{eqnarray}
&& P_{\alpha\beta} = P_{\alpha\beta}(L,E) = \nonumber \\
&& \delta_{\alpha\beta}
-4\sum_{a=1}^n\sum_{\beta=1, a<b}^n{\rm Re}\left(U_{\alpha a}^*
U_{\beta a}U_{\alpha b}U_{\beta b}^*\right){\rm sin}^2\left(\frac{\Delta m_{ab}^2 L}{4E}\right) - \nonumber \\
&& 2 \sum_{a=1}^n \sum_{b=1, a<b}^n {\rm Im}\left(U_{\alpha a}^*
U_{\beta a}U_{\alpha b}U_{\beta b}^*\right){\rm sin}\left(\frac{\Delta m_{ab}^2 L}{2E}\right) \label{oscprob}
\end{eqnarray}
where $\alpha, \beta = e,\mu, \tau, ...$, $a,b = 1,2,...n$,
$\Delta m_{ab}^2 = m_a^2 - m_b^2$.

In general there are uncertainties in the energy $E$
in the production of a $\nu$ (and/or ${\overline \nu})$-wave,
and also in the oscillation length.
As a result of these uncertainties
one has to
average the oscillation probability (\ref{oscprob})
over the $L/E$ dependence.

Considering a Gaussian average~\cite{ohlsson},
$$\langle P \rangle = \int_{-\infty}^{\infty} {\rm d}x~P(x)\frac{1}{\sigma \sqrt{2\pi}}
e^{-\frac{(x-\ell)^2}{2\sigma^2}}$$
$\ell \equiv \langle x \rangle$, $\sigma = \sqrt{\langle (x - \langle x \rangle)^2}$, $x = L/4E$,
and approximating
$\langle L/E\rangle \simeq \langle L \rangle /\langle E \rangle $
we obtain
\begin{eqnarray}
&& \langle P_{\alpha\beta} \rangle = \delta_{\alpha\beta}
- \nonumber \\
&& 2 \sum_{a=1}^n\sum_{\beta=1, a<b}^n{\rm Re}\left(U_{\alpha a}^*
U_{\beta a}U_{\alpha b}U_{\beta b}^*\right)
\left( 1 - {\rm cos}(2\ell \Delta m_{ab}^2)
e^{-2\sigma^2(\Delta m_{ab}^2)^2}\right)
\nonumber \\
&& -2 \sum_{a=1}^n \sum_{b=1, a<b}^n {\rm Im}\left(U_{\alpha a}^*
U_{\beta a}U_{\alpha b}U_{\beta b}^*\right)
{\rm sin}(2\ell \Delta m_{ab}^2)
e^{-2\sigma^2(\Delta m_{ab}^2)^2}
\label{average}
\end{eqnarray}
Notice the exponential
damping factors due to the fluctuations $\sigma$.

In fact, as discussed in \cite{ohlsson},
there are two kinds of bounds for $\sigma$:
A {\it Pessimistic:} one, according to which
$\sigma \simeq \Delta x \simeq \Delta\frac{L}{4E}
\le \frac{\langle L \rangle}{4 \langle E \rangle}
\left(\frac{\Delta L}{\langle L \rangle} + \frac{\Delta E}{\langle E \rangle}\right) $
and an {\it Optimistic:} $\sigma
\le \frac{\langle L \rangle}{4 \langle E \rangle}
\left([\frac{\Delta L}{\langle L \rangle}]^2 + [\frac{\Delta E}{\langle E \rangle}]^2\right)^{1/2}$.

In our case, where we consider long baseline experiments, the uncertainties
in the oscillation length $L$ are negligible, and hence the two cases degenerate
to a single expression for $\sigma =
\frac{\langle L \rangle}{4 \langle E \rangle}\frac{\Delta E}{E}~.$

The damping exponent, then, in (\ref{average}),
arising from the uncertainties
in the energy of the (anti)neutrino beam, becomes
\begin{equation}
2\sigma^2 (\Delta m^2)^2 = 2\frac{\langle L \rangle^2}{(4 \langle E
\rangle)^2} \left(\frac{\Delta E}{E}\right)^2(\Delta m^2)^2 ~.
\end{equation}
As mentioned above, for oscillation lengths
we have $L \Delta m^2/2E \sim {\cal O}(1)$,
and hence, the result of the best fit
(\ref{special}),(\ref{ceigen}), implying independence of the damping exponent
on $L$ (irrespective of the power of $L$), yields an uncertainty
in energy of order
\ba
 \frac{\Delta E}{E} \sim 1.6 \cdot 10^{-1}
\ea
if one assumes that this is the principal reason for the
result of the fit. This
is not an unreasonable result, thus implying that
the result of the fit may be interpreted as being due to ordinary
physics associated with uncertainties in the energy of the neutrino beam.

The important  difference from the stochastic fluctuations of gravity
medium, discussed above, lies on the fact that the period of oscillation is
not affected (\ref{average}) by the above averaging procedure,
in contrast to the stochastic gravity cases
(\ref{3genprob}) and
(\ref{gravstoch}), and thus in principle the effects can be disentangled.
However, in general these latter
corrections are small, and beyond the sensitivity
of the current experiments. Nevertheless, as we have seen,
some effects, such as the time-reversal symmetric
stochastic fluctuations
of the background metric (\ref{gravstoch}),
can be already
disentangled straightforwardly by their order as compared with the
above energy-uncertainty effects due to ordinary physics.

The precise energy and length dependence of the damping factors
is an essential step in order to determine the microscopic
origin of the induced decoherence and disentangle
genuine new physics effects~\cite{poland} from conventional
effects, which as we have seen above
may also contribute to decoherence-like damping~\cite{ohlsson2}.
For instance, as we discussed above,
some genuine quantum-gravity
effects, such as the stochastic fluctuations of the space time,
are expected to increase in general with the energy
of the probe~\cite{poland},
as a result of back reaction effects on space-time geometry,
in contrast to ordinary-matter-induced `fake' CPT violation and
`decoherence-looking' effects, which decrease with
the energy of the probe~\cite{ohlsson2}.
At present, the sensitivity of the experiments is not sufficient to
unambiguously determine the microscopic origin of the decoherence
effects, as we have seen above,
but we think that in the near future, when
experiments involving both higher energy
and precision become available, one would be able to
arrive at definite conclusions on this important issue.
Thus, phenomenological analyses like ours are of value and should be
actively pursued, in our opinion, in the future, not only in neutrino
physics but also in other sensitive probes of quantum mechanics, such
as neutral mesons.

\section{Conclusions and Outlook}

In this work we have presented a complete analysis of
three-generation  neutrino transition probabilities,
which include decoherence effects with
guaranteed positivity. In our opinion
this is the first complete, mathematically consistent, example
of a Lindblad-decoherence
model for neutrinos with full three generation mixing.

We have shown that decoherence
effects can account for  all available neutrino data,
including LSND results even in a minimalistic scenario (with only one new
parameter, which parametrises all the decoherence effects).
Contrary to other approaches in the literature,
using sterile neutrinos~\cite{mohapatra},
and following
the spirit of our earlier work~\cite{barenboim2,barenboim}, we have
attempted to interpret the LSND results not by means of oscillations, but
as a decoherence effect inducing damping in the oscillatory terms,
which is also present (as a per-mil-ish
additional suppression) in other neutrino experiments as well.

The specific oscillation length $L$ dependence of the single
decoherence parameter, 
implying $L$-independence of the corresponding damping exponents,
could in principle find 
a natural explanation in some theoretical
models of stochastic quantum gravity. However, its order of magnitude
seems incompatible with this possibility, at least in the concrete 
space-time foam microscopic models considered here, since it would imply
quantum-gravity effects unrealistically large.

On the other hand, the result of 
our fit can find a natural explanation in terms of 
ordinary physics. It could be due, for instance,
to uncertainties in the energy beam of (anti)neutrinos, 
and indeed this scenario seems to provide the most natural 
explanation of our fit. 

We now remark that quantum-gravity contributions
could indeed be present, and lead to similar damping 
in oscillatory terms, but their suppressed
order of magnitude would imply that they could only be probed
at  higher energies. For instance, 
high-energy neutrinos detected
from distant supernovae, may
probe these issues further, since they will increase significantly
the sensitivity to genuine quantum gravity effects~\cite{morgan}, and thus
may probe the induced changes in the damping as well as 
the oscillation period, as discussed  in this work.

It goes without saying that there is 
much more work to be done, both theoretical and
experimental, before definite conclusions are reached on this important
issue, but we believe that neutrino (astro)physics will provide a
very sensitive probe of new physics, including quantum gravity, in the
not-so-distant future.

\section*{Acknowledgements}
G.B. is partially supported by the spanish MEC and FEDER under FPA2005-01678 and
by Generalitat Valenciana under GV05/267.
The work of N.E.M. is partially supported by funds made available
by the European Social Fund (75\%) and National
Resources (25\%) - EPEAEK B - PYTHAGORAS (Greece).
G.B. and N.E.M. would like to thank M. Baldo-Ceolin for the
invitation to the {\it III International Workshop on Neutrino Oscillations
in Venice}, 7-10 February 2006, Venice (Italy), where preliminary
results of this work have been presented (N.E.M).


\begin{thebibliography}{99}

\bibitem{wheeler} See for instance:
  J.~A.~Wheeler and K.~Ford,
{\it Geons, black holes, and quantum foam: A life in physics,} and references therein.
\href{http://www.slac.stanford.edu/spires/find/hep/www?irn=4030974}{SPIRES entry}

\bibitem{poland} N.~E.~Mavromatos,
Lect.\ Notes Phys.\  {\bf 669}, 245 (2005)
[arXiv:gr-qc/0407005] and references therein.


\bibitem{ohlsson}
  T.~Ohlsson,
  Phys.\ Lett.\ B {\bf 502}, 159 (2001)
  [arXiv:hep-ph/0012272].




\bibitem{ohlsson2} M.~Blennow, T.~Ohlsson and W.~Winter,
  JHEP {\bf 0506}, 049 (2005)
  [arXiv:hep-ph/0502147];
M.~Jacobson and T.~Ohlsson,
  Phys.\ Rev.\ D {\bf 69}, 013003 (2004)
  [arXiv:hep-ph/0305064].

\bibitem{lindblad} G.~Lindblad,
  Commun.\ Math.\ Phys.\  {\bf 48}, 119 (1976);
R.~Alicki and K.~Lendi, Lect. Notes Phys. {\bf 286} (Speinger Verlag, Berlin
(1987)).


\bibitem{barenboim2} G.~Barenboim and N.~E.~Mavromatos,
  JHEP {\bf 0501}, 034 (2005)
  [arXiv:hep-ph/0404014].

\bibitem{lsnd}
  A.~Aguilar {\it et al.}  [LSND Collaboration],
  Phys.\ Rev.\ D {\bf 64}, 112007 (2001)
  [arXiv:hep-ex/0104049];
  G.~Drexlin,
  Nucl.\ Phys.\ Proc.\ Suppl.\  {\bf 118}, 146 (2003).





\bibitem{lisi} E.~Lisi, A.~Marrone and D.~Montanino,
  Phys.\ Rev.\ Lett.\  {\bf 85}, 1166 (2000)
  [arXiv:hep-ph/0002053].




\bibitem{gago}
  A.~M.~Gago, E.~M.~Santos, W.~J.~C.~Teves and R.~Zukanovich Funchal,
  arXiv:hep-ph/0208166.




\bibitem{lopez} J.~R.~Ellis, J.~S.~Hagelin, D.~V.~Nanopoulos and M.~Srednicki,
  Nucl.\ Phys.\ B {\bf 241}, 381 (1984);
J.~R.~Ellis, N.~E.~Mavromatos and D.~V.~Nanopoulos,
  Phys.\ Lett.\ B {\bf 293}, 142 (1992)
  [arXiv:hep-ph/9207268];
J.~R.~Ellis, J.~L.~Lopez, N.~E.~Mavromatos and D.~V.~Nanopoulos,
  Phys.\ Rev.\ D {\bf 53}, 3846 (1996)
  [arXiv:hep-ph/9505340];
P.~Huet and M.~E.~Peskin,
  Nucl.\ Phys.\ B {\bf 434}, 3 (1995)
  [arXiv:hep-ph/9403257].
F.~Benatti and R.~Floreanini,
  Phys.\ Lett.\ B {\bf 468}, 287 (1999)
  [arXiv:hep-ph/9910508];
  Nucl.\ Phys.\ B {\bf 511}, 550 (1998)
  [arXiv:hep-ph/9711240].



\bibitem{kamland} T.~Araki {\it et al.}  [KamLAND Collaboration],
  Phys.\ Rev.\ Lett.\  {\bf 94}, 081801 (2005)
  [arXiv:hep-ex/0406035].

\bibitem{bl} G.~Barenboim, L.~Borissov, J.~Lykken and A.~Y.~Smirnov,
JHEP {\bf 0210}, 001 (2002)
[arXiv:hep-ph/0108199].


\bibitem{msw2} N.~E.~Mavromatos, Sarben~Sarkar, to appear.


\bibitem{gorini}
    V.~Gorini, A.~Kossakowski and E.~C.~G.~Sudarshan,
    J.\ Math.\ Phys.\ {\bf 17}, 821 (1976).


\bibitem{loreti}
  F.~N.~Loreti and A.~B.~Balantekin,
  Phys.\ Rev.\ D {\bf 50}, 4762 (1994)
  [arXiv:nucl-th/9406003];
  E.~Torrente-Lujan,
  arXiv:hep-ph/0210037;
F.~Benatti and R.~Floreanini,
  Phys.\ Rev.\ D {\bf 71}, 013003 (2005)
  [arXiv:hep-ph/0412311].

\bibitem{mohapatra} see for instance:
A.~Strumia,
  Phys.\ Lett.\ B {\bf 539}, 91 (2002)
  [arXiv:hep-ph/0201134];
V.~Barger, D.~Marfatia and K.~Whisnant,
  Phys.\ Lett.\ B {\bf 576}, 303 (2003)
  [arXiv:hep-ph/0308299];
S.~Palomares-Ruiz,
  arXiv:hep-ph/0602083;
S.~Palomares-Ruiz, S.~Pascoli and T.~Schwetz,
  JHEP {\bf 0509}, 048 (2005)
  [arXiv:hep-ph/0505216].
For recent concise reviews on neutrino theories
see: R.~N.~Mohapatra {\it et al.},
  arXiv:hep-ph/0510213, and references therein.


\bibitem{barenboim} G.~Barenboim and N.~E.~Mavromatos,
  Phys.\ Rev.\ D {\bf 70}, 093015 (2004)
  [arXiv:hep-ph/0406035].



\bibitem{msw}
  L.~Wolfenstein,
  Phys.\ Rev.\ D {\bf 17}, 2369 (1978);
  S.~P.~Mikheev and A.~Y.~Smirnov,
  Sov.\ J.\ Nucl.\ Phys.\  {\bf 42}, 913 (1985)
  [Yad.\ Fiz.\  {\bf 42}, 1441 (1985)].


\bibitem{wald} R.~M.~Wald,
  Phys.\ Rev.\ D {\bf 21}, 2742 (1980).



\bibitem{morgan} D.~Hooper, D.~Morgan and E.~Winstanley,
  Phys.\ Lett.\ B {\bf 609}, 206 (2005)
  [arXiv:hep-ph/0410094];
L.~Anchordoqui, T.~Han, D.~Hooper and S.~Sarkar,
  Astropart.\ Phys.\  {\bf 25}, 14 (2006)
  [arXiv:hep-ph/0508312].

\end{thebibliography}
\end{document}